
\magnification=1200
\font\small=cmr8

 1
\font\bold=cmbx10
 1

\font\Bold=cmbx10 scaled\magstep 1
\font\chapter=cmbx10 scaled\magstep 2
 5
 3
\def\mpc{\,h^{-1}{\rm Mpc}}
\def\kms{\,{\rm km s^{-1}}}

\def\mpc{\,h^{-1}{\rm Mpc}}

\def\bx{{\bf x}}

\def\bxi{{\bar \xi}}

\def\bxiab{{\bar \xi}_{\rm AB}}
\def\xiab{ { \xi}_{\rm AB}}
\def\Ppab {P_{\rm AB}}
\def\Ppar {P_{\rm AR}}
\def\Piab {\Pi_{\rm AB}}
\def\Piar {\Pi_{\rm AR}}

%
\def\ref{\parskip=0pt\par\noindent\hangindent\parindent
    \parskip =2ex plus .5ex minus .1ex}
\def\gs{\mathrel{\raise1.16pt\hbox{$>$}\kern-7.0pt
\lower3.06pt\hbox{{$\scriptstyle \sim$}}}}
\def\ls{\mathrel{\raise1.16pt\hbox{$<$}\kern-7.0pt
\lower3.06pt\hbox{{$\scriptstyle \sim$}}}}
\def\gtsima{$\; \buildrel > \over \sim \;$}
\def\ltsima{$\; \buildrel < \over \sim \;$}
\def\prosima{$\; \buildrel \propto \over \sim \;$}
\def\gsim{\lower.5ex\hbox{\gtsima}}
\def\lsim{\lower.5ex\hbox{\ltsima}}
\def\simgt{\lower.5ex\hbox{\gtsima}}
\def\simlt{\lower.5ex\hbox{\ltsima}}
\def\simpr{\lower.5ex\hbox{\prosima}}
%
%
%
\vskip3truecm

\baselineskip=12pt
\font\small=cmr8

\topskip=8.5truecm
\centerline{\Bold SPATIAL DISTRIBUTION}
\smallskip
\centerline {\Bold OF LOW SURFACE BRIGHTNESS GALAXIES}
\vskip20pt
\centerline {\rm H.J. Mo$^1$, Stacy S. McGaugh$^1$, Gregory D. Bothun$^2$}
\vskip10pt
\centerline {\small 1 Institute of Astronomy,
Madingley Road, Cambridge CB3 0HA, UK}
\vskip10pt
\centerline {\small 2 Department of Physics, University of Oregon,
Eugene OR 97403, USA}
\vskip130pt
\centerline {submitted to {MNRAS}}
\bigskip
\bigskip
\vfill
\eject

\topskip=0.0truecm
\centerline{\bf SPATIAL DISTRIBUTION}
\smallskip
\centerline {\bf OF LOW SURFACE BRIGHTNESS GALAXIES}
\vskip15pt
\centerline {\rm H.J. Mo$^1$, Stacy S. McGaugh$^1$, Gregory D. Bothun$^2$}
\vskip10pt
\centerline {\small 1 Institute of Astronomy,
Madingley Road, Cambridge CB3 0HA, UK}
\vskip10pt
\centerline {\small 2 Department of Physics, University of Oregon,
Eugene OR 97403, USA}
\vskip15pt
\centerline {submitted to {MNRAS}}

\baselineskip=15pt
\centerline{\Bold Abstract}
\smallskip

The spatial distribution of low surface brightness (LSB)
galaxies is important both
as a test of theories of large scale structure formation and to
the physical understanding of the environmental effects which influence the
evolution of galaxies.  In this paper we calculate,
using redshift samples, the cross-correlation functions
[$\xiab (r)$] of LSB galaxies with (normal galaxies) in complete samples
(i.e. CfA and IRAS), which enables us to
compare directly the amplitudes and shapes of the correlation
functions for LSB galaxies to those for CfA and IRAS galaxies.
For pair separations $r\gs 2\mpc$, we find $\xiab (r)\propto
r^{-\gamma}$ with $\gamma\approx 1.7$. This shape of
$\xiab$ is in agreement with that of the correlation functions for
other galaxies. The amplitudes ($A$) of $\xiab (r)$ are lower
than those of the autocorrelation functions for the CfA and IRAS
samples, with
$A_{\rm LSB-CfA}:A_{\rm CfA-CfA}\approx 0.4$ and
$A_{\rm LSB-IRAS}:A_{\rm IRAS-IRAS}\approx 0.6$.
These results suggest that LSB galaxies are imbedded in the same
large scale structure as other galaxies, but are less strongly clustered.
This offers the hope that LSB galaxies may be unbiased tracers
of the mass density on large scales.
For $r\ls 2\mpc$, the cross-correlation functions
are significantly lower than that expected from the extrapolation
of $\xiab$ on larger scales, showing that the formation and survival
of LSB galaxies may be inhibited by interaction
with neighboring galaxies.

We show that a simple hierarchical model, in which LSB
galaxies are formed only in halos lacking close interactions
with other halos,
reproduces both the deficit of
pairs at small separations and the
low amplitude of the correlation function.
This model suggests that LSB galaxies should, on average,
be younger than normal galaxies, consistent with direct observations.
The model also suggests that a strong luminosity (mass) segregation
in galaxy clustering is not a necessary consequence of biased
galaxy formation, unless the effect of surface brightness (collapse
time) is taken in to account.
It is also possible that a significant fraction of the mass density
of the universe resides in galaxies which
have not been observed because of their low surface brightness.

\smallskip
\leftline {{\bf Key words}: galaxies: clustering -- galaxies:
formation -- cosmology: observations.}

\vfill\eject

\baselineskip=20pt
\bigskip
\centerline{\Bold 1. Introduction}
\bigskip

According to the theory of biased galaxy formation in an Omega = 1
Universe
(see e.g.
Kaiser 1984; Bardeen et al. 1986, hereafter BBKS),
normal and easily detectable high surface brightness (HSB)
are more strongly clustered than the underlying mass distribution.
However, strong selection effects act against the
discovery of galaxies of low surface brightness (LSB) (Disney 1976).  If
LSB galaxies trace the mass distribution in a more representative
manner than HSB objects then clearly our perception of galaxy clustering
is biased as well.  Indeed,
it has been suggested that LSB dwarfs corresponding
to $\sim 1\sigma$ fluctuations in the initial density field
should trace mass and fill the voids (e.g.\ Dekel \& Silk 1986).

However, not all LSB galaxies are dwarfs (Bothun et al. 1987,
Davies et al. 1988, Impey \& Bothun 1989,
Irwin et al. 1990, Bothun et al. 1990, McGaugh 1992, Knezek 1992, McGaugh \&
Bothun 1993, van der Hulst \& de Blok 1993).  Indeed, the most
common types of objects found in modern field surveys are comparable
in size and mass to HSB spirals (Schombert \& Bothun 1988,
Schombert et al. 1992, Impey et al. 1993).  These objects apparently
do not fill the voids, and seem to trace the same large scale structures
seen in the same part of the sky by the CfA redshift survey (Schombert
et al. 1992).
Nonetheless, the low densities of LSB galaxies do
suggest that they are associated with formation from relatively
low peaks in the initial density field (McGaugh et al. 1993),
albeit on the mass scale of spiral rather than dwarf galaxies.
In this case they trace mass in a different and perhaps more
representative way than HSB galaxies (c.f.\ Davis \& Djorgovski 1985).

Recently it has been noted that LSB disk galaxies
have a highly significant deficit of companions within $\sim 1 \mpc$
(Zaritsky and Lorrimer 1992, Bothun et al. 1993), even though there
is no obvious segregation by
surface brightness on larger scales (Bothun et al. 1986,
Thuan et al. 1987, Schneider et al. 1990).
This effect seems to be especially pronounced for giant disks like Malin~1
(Impey and Bothun 1989, Bothun et al. 1990, Knezek 1992), which appear to be
the sole massive occupants of large scale underdense regions.
Indeed, Malin 1 exhibits
the same global properties (in terms of size and mass)
as a small group of galaxies yet only one object
actually formed.

In this paper we examine the correlation properties
of LSB galaxies
by utilizing a technique which is insensitive to the
selection effects against the objects in question.  We explicitly compare
the correlation properties of a sample of LSB galaxies to those of
a nearly complete
sample of HSB galaxies and to IRAS galaxies.  This comparison is now possible
due to the ongoing surveys for LSB objects (e.g.\ Schombert et al. 1992,
Impey et al. 1993) that have yielded redshifts for a few hundred LSB disk
galaxies.

\bigskip
\centerline {\Bold 2. The data}
\bigskip

In order to study the distribution of LSB galaxies,
samples of which are by nature incomplete,
we need galaxies (called tracers) which sample
the galaxy density field.  We
use two complete samples of galaxies as tracers:
the CfA redshift survey of galaxies (Huchra
et al. 1983); and the redshift survey of
IRAS galaxies brighter than 2 Jy
(Strauss et al. 1992 and references therein).
The CfA survey of optical galaxies
is complete to $m=14.5$ and contains about 2400 galaxies
in the regions
($\delta \ge 0^\circ, b^{\rm II}\ge 40^\circ)$ and
($\delta \ge -2.5^\circ, b^{\rm II}\le -30^\circ)$.
The IRAS sample,
a redshift survey of galaxies detected by the
{\it Infrared Astronomical Satellite},
includes  2652 galaxies over 87.6\% of the sky
(with $\vert b^{\rm II} \vert >5^\circ$ excluding small
areas masked out in the survey)
with 60 $\mu$ flux $f_{60}>1.936$Jy.
Figures 1 and 2 display, respectively,
the sky and redshift
distributions of these samples.
A comparison of these distributions with
those of the LSB sample (summarized below)
indicates a good overlap in space.
To normalize the correlation functions we need the
selection functions which give the relative
probability that a trace galaxy is included in the
trace samples.
For the CfA sample, we use the selection function
given by Strauss, Yahil \& Davis (1991).
For the IRAS sample we use that given by Yahil et al. (1991).
These two selection functions are also plotted
in Figure 2 (the upper panel).
It should be pointed out that, at
our present stage of
understanding of galaxy formation, it is not clear
whether either
population of galaxies
is the appropriate `tracer' of the galaxy density field;
it may well be possible that the number density of galaxies
is dominated by a unknown population of galaxies.
The galaxy density field we will discuss
applies only to that given by the observed optical
and IRAS galaxies.

For the targets representing the distribution of LSB
galaxies, we use a sample constructed from
Schombert et al. (1992).
This sample (called POSS in the following)
consists of 339 galaxies (171 with redshifts)
discovered by visual inspection of the plates
of the Second Palomar Sky Survey.
Like the UGC (Nilson 1973), these objects are selected by a
diameter limit, and so
probe the range of surface brightness between the
isophotal limits of the old (25.3\sb; Cornell et al. 1987)
and new (26.0\sb; Schombert \& Bothun 1988) sky survey plates.
In general, these newly discovered galaxies are very difficult to
discern on the old POSS plates and hence it is not suprising
that they were passed over by Nilson.
The new POSS sample thus defines a slice in surface brightness
beyond the UGC (which already contains some quite low surface
brightness galaxies) which is well removed from the high surface
brightness objects typical of the CfA sample.
This provides a much cleaner signal in the correlation analysis
than can be attained by simply splitting up the UGC (e.g.\ Davis \&
Djorgovski 1985, Bothun et al. 1986).

The sky distribution of
POSS galaxies with measured redshifts is shown in
the middle panel of Fig.1. The redshift distribution
of these galaxies is shown in Fig.2.
These LSB galaxies are virtually never detected
by IRAS (Schombert \& Bothun 1988), so there is no
concern about coincidence between the trace and target
galaxies in this case.  Coincidence with the CfA sample
has been discussed in detail by Bothun et al. (1993).

\bigskip
\bigskip
\centerline {\Bold 3. Statistical method}
\bigskip

In order to study the distribution of LSB galaxies
in the galaxy density field,
we can calculate the
mean density of trace galaxies interior to a sphere
of radius $r$, around each target (LSB) galaxy.
Suppose the two-point cross-correlation function between
the target galaxies of type A with trace galaxies B
is $\xi_{\rm AB}(r)$ (which is defined as the excess probability
of finding a trace galaxy B in a
spherical shell of radius $r$ centered
on a randomly chosen target galaxy of type A;
see Peebles 1980 \S44).  The mean density of
the trace galaxies inside the sphere
is
$$ n(r)=n_{\rm B}[1+\bxiab(r)]\eqno(1)$$
where
$$
\bxiab(r)={3\over r^3}\int _0^r \xiab (x) x^2 {\rm d} x ,
\eqno(2)
$$
and $n_{\rm B}$ is the overall mean density of trace galaxies.
This analysis technique has been widely used
to study the relative distributions
of two populations (see Mo \& Lahav
1993 for a discussion).
It is not essential that the target sample be complete
(actually one can calculate a cross-correlation function for
each target galaxy),
provided that the target sample is a fair sample and its incompleteness
does not correlate with galaxy clustering.
This advantage is important to our analysis because
it is virtually impossible to to construct a sample
which is complete in LSB galaxies.  Regardless of the
depth of the isophotal limit of a survey, there may always
exist objects of such low surface brightness that their apparent
diameters do not exceed the survey limit
(Disney 1976).  While this selection against low contrast objects
is strong, it is an observational limitation which should in no
way be correlated with the spatial distribution of galaxies.
That is, there is no evidence that either Nilson (1973)
or Schombert \& Bothun (1988) and Schombert et al. (1992)
looked harder to find LSB objects in underdense regions.
The sky coverage of the target sample is fairly small; it is not known
whether the clustering of LSB galaxies in this region is similar to that
of a fair sample of the universe. However, as shown in section 4,
the clustering of HSB galaxies in this region is similar to what
we known from other samples.

We estimate the cross correlation function
$\xiab (r)$ using the definition
$$\xiab(r)={\Ppab (r)\over \Ppar (r)}-1, \eqno(3)$$
where $\Ppab (r)$ is the number of cross pairs between
target objects (A) and trace objects (B),
with separation $r$, in the data;
$\Ppar $ is the corresponding number of cross pairs
between target objects (A) and random points (R)
in a random sample which has the same selection
function as the trace sample.
The average function $\bxiab (r)$ can either be
calculated by using Eq. (2), or estimated
from the definition,
$$\bxiab(r)={\Piab (r)\over \Piar (r)}-1 , \eqno(4)$$
Here, $\Piab (r)$ is the number of cross pairs between
target objects (A) and trace objects (B),
with separation $\le r$, in the data;
$\Piar (r)$ is the corresponding expected number
of pairs.
For our cases,
these two definitions of
$\bxiab$ give very similar results.
The advantage of the second definition is that
it is more stable and independent of the choice of
bins in $r$ when $r$ is large.

The distances of galaxies are calculated
from their redshifts, so these are
correlation functions in {\it redshift space}.
The redshifts are corrected
for a velocity of 300 km/s due to the motion
in the Local Group (LG), and for the Virgocentric
flow in the way described
by Schechter (1980). The model assumes
a Virgocentric infall velocity field varying
inversely as the Virgocentric distance and
presumes the mean (LG-frame) velocity of the
Virgo cluster to be 1020 km/s.
The Virgocentric infall of the LG is assumed
to be $v_{\rm LG}=220$ km/s.
The median redshift of the LSB sample is over 5000 km/s,
so our results should not be sensitive to the details
of local distortions in the Hubble flow.  To test this,
we have also used infall velocities of
$v_{\rm LG}=100$ and 300 km/s
to bracket the realistic values. Our results do not change
significantly for these values of $v_{\rm LG}$.
Galaxies within $6^\circ$ of the
center of Virgo and with measured velocities less
than 2500 km/s are assumed to be at the mean
Virgo velocity.

As we can see from Eqs. (3) and (4),
an accurate estimate of the cross
correlation functions depends on an accurate estimate
of the expected number of pairs $\Ppar$
(or $\Piar$) which,
in turn, depends on our understanding of the sample
selection effects. In our case, an important advantage is that
the selection functions of both CfA and IRAS samples
are reasonably well known.
Since our trace samples are apparent-magnitude
or flux limited, the pair counts will be
dominated by the contributions from the nearby region
if trace galaxies are not weighted by their
selection functions.
In most cases, we calculate the cross-correlation functions
by assigning to each trace galaxy a weight
which is proportional to the reciprocal
of the selection function at the distance of the
galaxy. Such a weighting scheme gives, however,
quite noisy results on small scales, because
there are not many close pairs
of galaxies at large distances.
To estimate the correlation function on small scales
(e.g. $r\ls 1\mpc$),
we use a scheme in which
all galaxies are equally weighted.
Such a change of weighting scheme on small scales improves
the statistics but does not alter the behaviour of the correlation
function.
The statistical uncertainty in correlation functions
is difficult to analyse.
To present our results, we will use
errors given by the bootstrap resamplings
of the target samples
(see e.g. Barrow, Bhavsar \& Sonoda 1984;
Mo, Jing \& B\"orner 1992).

\bigskip
\bigskip
\centerline {\Bold 4. Results}
\bigskip

We calculate the cross correlation functions
for the various cases summarized in Table 1.
The first column of Table 1 lists the samples which
are used to calculate the cross correlation functions.
A case denoted by `Sample1-Sample2' means, for example,
that `Sample1' is to be used as the target sample while `Sample2'
is the trace sample. The second column indicates
the sky region to which the {\it target} galaxies are
confined.  For all cases, we exclude
galaxies with corrected redshifts larger
than 10000 km/s.
For the CfA sample, we also exclude galaxies
with absolute magnitudes fainter than $-18.5$.
This is done to reduce the weight of local
superclusters. This restriction should not influence the results
on larger scales as the correlation function does not depend
strongly on luminosity (Phillipps \& Shanks 1987; Hamilton 1988;
Alimi, Valls-Gabaud, \& Blanchard 1988; B\"orner \& Mo 1990).
The number of galaxies in the
target sample is listed in the third column for
each case.

Figure 3 shows the average correlation functions
($\bxiab$)
for the cases CfA-CfA (triangles)
and POSS-CfA (bullets).
The correlation functions are normalized by a random CfA catalogue
generated by the selection function (SF) given by Strauss et al. (1991).
This SF is in good agreement with that derived from the luminosity
function (of the same sample) given by Efstathiou, Ellis \& Peterson (1988).
The straight line corresponds to a differential
correlation function, $\xi (r)=(6 \mpc /r)^{1.7}$,
which is approximately the result of the {\it redshift
space} auto-correlation function obtained by Davis \& Peebles
(1983) for the CfA sample. That this line fits reasonably
well the result of our CfA-CfA analysis demonstrates that the POSS
region is not a peculiar region of galaxy clustering.
The error bars show the 1 sigma standard deviation
among the results of 100 bootstrap resamplings of the
POSS sample. Clearly, the cross
correlation function for POSS-CfA has a systematically lower
amplitude, and much flatter shape on small scales,
than that for CfA-CfA.  Figure 4 shows the cross
correlation function in its differential form
(Eq.3).

Since it is known that spiral galaxies
are less strongly correlated than elliptical
galaxies (Davis \& Geller 1976;
Giovanelli et al. 1986; B\"orner, Mo \& Zhou 1989;
Jing, Mo \& B\"orner 1991; Mo et al. 1992),
and that the LSB galaxies in our samples are disk galaxies,
it is interesting to see what happens when
spiral galaxies alone are used to trace the galaxy density
field. Figure 5 shows the results when all
galaxies with types later than $T=0$
in the CfA sample are used (this sample
is called `Spiral' in our discussion).
In this case, we have used the selection function
for the total sample, based on the result of Davis
and Huchra (1982) that the shape of the luminosity function
does not depend significantly on morphological type.
This SF is almost identical to that derived from the luminosity
function for all spiral galaxies (with $T>1$) given by Efstathiou
et al. (1988a). For comparison, we also show in Fig.5 the LSB-spiral
cross-correlation function estimated by using a SF derived from the luminosity
function of all spiral galaxies given by Davis \& Huchra (1982).
Since these two SFs differ substantially from each other, the comparison shows
that our results are not sensitive to the SF used.
The amplitude of the correlation function
for Spiral-Spiral is about 0.7 times that
for the total sample, in good agreement with
that found by B\"orner et al. (1989) and
Jing et al. (1991).
The cross correlation
function for POSS-Spiral has, however, a lower
amplitude and a flatter shape than
that for Spiral-Spiral,
in a similar manner as that shown in Fig.3.

Figure 6 shows the average correlation functions
($\bxiab$)
for the cases IRAS-IRAS (triangles)
and POSS-IRAS (bullets).
The correlation functions are normalized by a random
IRAS catalogue generated by using the SF given by Yahil
et al. (1991).
The straight line corresponds to a differential
correlation function, $\xi (r)=(4 \mpc /r)^{1.6}$,
which is approximately the result for the
{\it redshift space}
auto-correlation function obtained by Strauss
et al. (1992) for the total IRAS sample.
Our correlation function for IRAS-IRAS is in agreement
with this result, showing again that the region
in consideration is not peculiar in galaxy clustering.
Figure 7 shows the correlation function in
differential form. The result
is noisy because of the low space density
of the IRAS sample.
In general, the results
show that the cross-correlation
function for POSS-IRAS has a
systematically lower
amplitude
and (marginally) an extra deficit at small separations.

\bigskip
\leftline {\bf 4.1. The amplitudes of the cross
correlation functions}
\bigskip

As we can see from Figs.1-7, the correlation functions
are well described by a power law for $r\gs 2\mpc$.
This enables us to compare the correlation amplitudes
for different cases. Such a comparison is important,
because it may give the relative bias factors for
the different galaxies in consideration, as we
will discuss in section 5.
We have performed a formal regression
for each of the results in
the range $2\mpc<r<10\mpc$ using a power
law model
$$
\bxiab (r)=A r^{-\gamma}
\eqno(5)
$$
where the amplitude $A$ and the power index
$\gamma$ are constants.
To compare the correlation amplitude more directly,
we have performed another regression by fixing
$\gamma$ to be 1.65, approximately the mean value
of $\gamma$ for the different cases.
The regression results are
presented in columns 4 and 5 in
Table 1.
For comparison we also list
(in column 6) the value
$r_0\equiv [A (3-\gamma)/3]^{1/\gamma}$,
which is the correlation length at which $\xi=1$ if the
power-law (5) is adopted.
The errors quoted are
derived by using the bootstrap errors for individual data points.
Since the errors of different data points may
not be independent, the real errors for both $\gamma$
and $A$ may be larger.

The values of the power index $\gamma$ and
the correlation length $r_0$ we obtained for
CfA-CfA are in good agreement with those obtained
by Davis \& Peebles (1983) for the total sample.
For IRAS-IRAS,
the power index $\gamma \sim 1.9$
we obtained
is much larger than the fiducial value
$\gamma =1.6$. This discrepancy is due to
the fact that there is a
shoulder around $r=4\mpc$ in our
correlation function, and that the regression
is performed for $r>2 \mpc$.
Using all data points in the range
$0.3\mpc <r < 10 \mpc$, we obtain
$\gamma =1.60$ and $r_0=3.5\mpc$.

 From Table 1, we can read off the following ratios
between the amplitudes of the correlation functions
(with $\gamma$ fixed to be 1.65):
$$A_{\rm C-C}:A_{\rm S-S}:A_{\rm I-I}
\approx 1:(0.74\pm 0.16):(0.48\pm 0.08) ; \eqno(6a)$$
$$A_{\rm L-C}:A_{\rm L-S}:A_{\rm L-I}
\approx 1:(0.77\pm0.20):(0.63\pm 0.14) ; \eqno(6b)$$
$$A_{\rm L-C}:A_{\rm C-C}\approx 0.44\pm 0.09 ; $$
$$A_{\rm L-S}:A_{\rm S-S}\approx 0.46\pm 0.12 ; \eqno(6c) $$
$$A_{\rm L-I}:A_{\rm I-I}\approx 0.57\pm 0.11 , $$
where C-C denotes CfA-CfA; S-S denotes Spiral-Spiral;
I-I denotes IRAS-IRAS, L-C denote LSB-CfA and so on.
The differences between the values in Eq.(6c) are
only marginally significant.
However, the basic result is clear:
the amplitudes of the cross-correlation functions between
LSB galaxies and trace galaxies are lower
than the corresponding amplitudes of the auto-correlation
functions of the trace galaxies
by a factor of about 2.

\bigskip
\leftline {\bf 4.2. The shapes of the cross
correlation functions}
\bigskip

A remarkable
property in the cross-correlation functions between
the LSB galaxies and the trace galaxies is the flattening
on small scales ($r\ls 2\mpc$).
Such a feature is absent in the auto-correlation functions
of the trace galaxies, and corresponds to the strong effect
on small scales reported by Bothun et al.
(1993). To show this feature, we
calculate the ratio, $ {\cal R} (r)$,
between the cross-correlation function
and the corresponding auto-correlation function of trace galaxies.
These ratios are shown in Figure 8.
Starting around $r\sim 2\mpc$ the ratios decrease
from the global values given by the correlation
amplitudes listed in Table 1 to
values that are
about five to ten times smaller
at $r\ls 1\mpc$.
In order to quantify this,
we have performed a formal
fit to the ratios,
using a model of the form
$${\cal R} (r)={\cal R}_0 \left\lbrack 1- {\rm exp}
\left\lbrace
-\left({r\over r_c}\right)^\beta\right\rbrace\right\rbrack.\eqno(7)$$
The coefficient ${\cal R}_0$ is fixed to be the values
given by the ratios between the amplitudes
of the correlation functions (Eq.6c).
The values of $\beta$ and $r_c$, given by a least-squares
fit, are listed in Table 1. The values of $r_c$, given by
a similar fit but with fixed $\beta = 2$, are also listed in Table 1.
The results of ${\cal R} (r)$ given by this model
(i.e. with $\beta=2$) are shown in Fig.8 by the solid curves.
The decrease of the ratios for small $r$ is
quite rapid, with $\beta=2$ giving a reasonable fit to the data.
The value of $r_c$ is about $2\mpc$ for the CfA case,
and about $1.0\mpc$ for the IRAS case, as can be seen by inspection
of Fig.8.

Our correlation functions
are estimated by using (corrected) galaxy redshifts
as distances. We should therefore consider how our
results are affected by having worked in
redshift rather than real space.
There are two effects:
the amplitudes of the correlation functions
are scaled on large scales, and the shapes are
altered on small scales. The basic principles
of redshift-space distortion were given by Kaiser (1987).
The first effect should be small, or even
absent in our cases, because we are considering fluctuations
of the galaxy density field on a scale between 2 to $10 \mpc$
where the redshift-space enhancement of the correlation function
is small in this regime (Suto \& Suginohara 1991).
The other effect of redshift-space distortions is
that virialized random peculiar velocities tend
to damp power at small wavelengths, which also
reduces $\xi$ at small separations.
The effect
can be modelled as
a Gaussian convolution
in the radial direction, and depends on the
peculiar velocity involved. For a typical
galaxy pairwise dispersion of 300 kms$^{-1}$, the
smearing `$\sigma$' is about $2\mpc$.
This would appear to be worrisome, because it
is comparable to the values of
$r_c$ which we derived for the flattening scales
of the cross-correlation functions between
LSB and CfA (or IRAS) galaxies.
However, if the smearing had such a big effect,
we would expect to see a similar or even
stronger effect in
the auto-correlation functions of CfA and IRAS
galaxies, because they are more strongly clustered
and presumably have larger
pairwise velocity dispersions. That such an effect
is not seen in the auto-correlation functions
suggests that the smearing does not have a significant
effect on their amplitudes. Indeed, if smearing is to be
significant, there must be a conspiracy such that the
real space correlation function changes on just the right
scale by just the right amount to maintain the smooth
power laws observed in redshift space.

As a check we have calculated the projected
cross-correlation function, $w(r_p)$
(where $r_p$ is the projected separation of galaxy
pairs), between the LSB galaxies
and CfA galaxies. The projected function is defined
and calculated in the same way as in Davis \& Peebles (1983).
The result is shown in Figure 9.
The dotted line has a slope of $-0.7$ which
corresponds to a {\it real space} power law
correlation function with slope $\gamma=1.7$.
As discussed by Davis \& Peebles (1983),
the projected function should not be affected
significantly by the peculiar velocities
of galaxies. Clearly,
the data have a much more flattened shape on small scales
than expected from the power law. Such a flattening
is not found in the auto-correlation functions
of CfA and IRAS samples (Mo, Jing, \& B\"orner 1993).
However, a more accurate estimate of $w(r_p)$ needs
samples with larger sky coverage and higher surface
density. We are unable to calculate $w(r_p)$
for the IRAS case because of the low surface
density of the sample.  Nonetheless, the behaviour
of $w(r_p)$ must be real and not a result of smearing
effects, or Bothun et al. (1993) would not have detected
the strong effect on small scales.

\bigskip
\centerline {\Bold 5. Discussion}
\bigskip

In this section we show how simple assumptions
about the formation of LSB galaxies
can reproduce the general properties of
the cross-correlation functions found in this paper.
We consider a simple hierarchical model in which
galaxies are associated with regions
identified in the initial density field.
These regions form halos within which the visible
galaxies form.  The halos are presumably dominated by
dark matter, but this is not essential to the model,
which follows from the identification of galaxies with
density fluctuations in the initial conditions and so depends only
on the properties of Gaussian random fields.  This
simple model must be considered no more than speculative
at the present level of our understanding of the physics
of galaxy formation, but does help to
illuminate the basic principles involved.

Within this framework, we identify LSB galaxies
with those halos which are sufficiently isolated that they
have not been incorporated into larger systems
by the present epoch.
This is motivated by two inferences drawn from
the physical properties of LSB galaxies.  First,
the slow evolution (McGaugh 1993) and young stellar
populations (McGaugh \& Bothun 1993) together
with the low star formation rates (McGaugh 1992)
and low gas surface densities (van~der~Hulst et al. 1993)
of LSB galaxies imply that they collapse late as the result
of formation from relatively low ($\simlt 1 \sigma$) peaks
in the initial density field (McGaugh et al. 1993).  Second
is the apparent lack of tidal interactions experienced by
LSB galaxies over a Hubble time (Zaritsky \& Lorrimer 1992,
Bothun et al. 1993) as evidenced by the
lack of nearby neighbors and star formation histories
incompatible with burst and fade scenarios (Schombert et al.
1990; McGaugh \& Bothun 1993).
Our model for the formation of LSB galaxies will satisfy both
of these, as galaxies which lack neighbours will not suffer tidal
interactions and will, on average, form from
fluctuations with lower initial amplitudes
(the inverse of the oft stated property of Gaussian fields
that high peaks occur preferentially in regions of high density).
This leads naturally
to a correlation function (for LSB galaxies) which
has a lower overall amplitude
and a more flattened shape on small scales than
that for average galaxies (average in the sense that
no restriction is made on the location of their halos).
Indeed, it will be seen that the low average height
of fluctuations
corresponds to the low amplitude of the correlation
function, and the need to avoid tidal interactions which enhance
star formation (and thus surface brightness) corresponds to the
flattening seen at small scales.

\bigskip
\leftline {\bold 5.1. Correlation functions}
\bigskip

To model the correlation properties of LSB galaxies,
we assume that the initial overdensity field,
$\delta (\bx)$, is Gaussian
and described by a power spectrum $P(k)$.
In particular, we consider CDM models
using the spectrum given by BBKS
This choice is made for specificity;
the general results are similar for any hierarchical scenario
with Gaussian initial conditions.

The probability of finding
a spherical region of comoving radius $R$ with mean
overdensity $\delta $ (linearly extrapolated to present time) is
$$
p(\nu ) d\nu  ={1\over \sqrt{2\pi}}
{\rm e}^{-\nu ^2 /2} d\nu ,
\eqno(8)
$$
where $\nu =\delta  /\Delta $;
$\Delta \equiv \Delta (R)$ is rms linear overdensity
in a sphere of radius $R$, extrapolated to present time.
For a spherical perturbation and $\Omega=1$,
each shell recollapses
to the origin at a time when the mean linear overdensity
interior to it extrapolates to the value 1.68
(see e.g., Peebles 1980). It is therefore assumed
that a region will be part of a
single collapsed structure (halo) at redshift $z$, if
its linear overdensity, extrapolated to the
present epoch, is $1.68 (1+z)$
(see e.g. Efstathiou et al. 1988b; White \& Frenk 1991).
The mass and circular velocity of a halo are,
assuming the physical overdensity in the virialized
sphere to be 200, related to
the comoving radius $R$ (in present units)
and redshift $z$ by
$$
M={4\pi \over 3} \rho _0R^3; \,\,\,
V_c=1.67 (1+z)^{1/2} H_0 R ,
\eqno(9)
$$
where $\rho_0$ and $H_0$ are, respectively,
the mean density of the Universe and the
Hubble constant at the present time.
So, the quantities
$M$, $V_c$, $\Delta$ and $R$ are
equivalent variables for a given redshift.

If we choose $R$ to be the linear scale
(in present units)
that gives $M$ the value of the
mass of typical galaxies (e.g. $R\sim 0.5\mpc$),
a region in the density field where
$\delta  > 1.68(1+z)$
will either itself be a halo for galaxy formation, or
be a part of a larger halo which may form a bigger galaxy
or a system of galaxies. To distinguish between these two
different cases requires a complete description
of the merging history of
halos. This is a complicated
problem and will not be considered here.
We will assume, instead, that once a region
in the density field (smoothed on galactic scales) reaches
the critical overdensity [$1.68(1+z)$]
at redshift $z$, a galactic halo
will form and exist in isolation or in a larger
system without loosing identification.
This is equivalent to assuming that the merger
of two or more galactic-size halos to form a single
galactic halo is not frequent.
Such an assumption
will not change our discussion of the correlation
function of LSB galaxies, because in our model
we assume that LSB galaxies are formed in halos
which are sufficiently isolated that they
have not been incorporated into larger collapsed
systems containing other galaxies by the present
epoch.

According to Bower (1991; see also Bond et al. 1991),
the fraction of the mass in the universe
that is in halos with
mass in the range $M \to M + d M$ at redshift
$z =\delta /\delta _c -1$, is
$$
f(\Delta , \delta )
d \Delta ^2 =
{1\over (2\pi)^{1/2}} {\delta
\over \Delta ^3}
{\rm exp} \left\lbrack -{\delta ^2
\over 2\Delta ^2 } \right\rbrack
d \Delta ^2 .
\eqno(10)
$$
Similarly,
the fraction of the mass $M_0$ of a spherical region with
(linearly extrapolated) overdensity $\delta _0$,
that is in halos with
mass in the range $M_1 \to M_1 + d M_1$
($M_1<M_0$) at redshift
$z_1 =\delta _1/\delta _c -1$, is
$$
f(\Delta _1, \delta _1\vert \Delta _0, \delta _0)
d \Delta _1^2 =
{1\over (2\pi)^{1/2}} {\delta _1-\delta _0
\over (\Delta _1^2 -\Delta _0^2)^{3/2}}
{\rm exp} \left\lbrack -{(\delta_1 -\delta _0)^2
\over 2(\Delta _1^2 -\Delta _0^2)} \right\rbrack
d \Delta _1^2 .
\eqno(11)
$$
So the number of $M_1$ haloes in such a spherical region is
$$
{\cal N}(1|0)\equiv {M_0\over M_1}
f(\Delta _1, \delta _1 | \Delta _0, \delta _0).
\eqno(12)
$$
The average two-point correlation function (in Lagrabgian space)
at redshift $z_0$, between mass and halos
with mass $M_1$ at redshift $z_1=\delta_1/\delta_c-1$
(called haloes of type 1),
$\bxi _{1m} (R_0)$, can be formally written as
$$
\bxi_{1m} (R_0)=\int \delta _0 p(0|1) d\delta_0,
\eqno(13)
$$
where $p(0|1)$ is the probability of finding a region with
(linear) radius $R_0$ and overdensity $\delta_0$, given that
there is in this region a halo of type 1. According to Bayes'
theorem, $p(0|1)\propto p(1|0)p(0)$, where $p(0)$ is the probability
of finding a region with radius $R_0$ and overdensity $\delta _0$;
$p(1|0)$ is the probability of finding a halo of type 1 in such a
region. So $p(1|0)\propto {\cal N} (1|0)$. The correlation function
$\bxi_{1m} (R_0)$ can now be written as
$$
\bxi_{1m} (R_0)={
\int _{-\infty} ^{1.68 (1+z_0)}
\delta _0 {\cal N}(1\vert 0) p(0) d \delta _0
\over
\int _{-\infty} ^{1.68(1+z_0)}
{\cal N} (1\vert 0) p(0)d \delta _0} ,
\eqno(14)
$$
where $p(0)$ is given by Eq.(8)
for a fixed $R_0$,
$z_0< z_1$, and the
integration limits are chosen so that the
spherical region of radius $R_0$ has not collapsed
by redshift $z_0$ (see Mo, Miralda-Escude and Rees 1993
for a discussion)

It should be noted that Eq.(14) gives only the
correlation function in comoving space, without
taking into account any dynamical evolution.
To treat the dynamical evolution accurately one needs
to invoke numerical simulations.
Here we calculate the correlation function in physical
space, using a simple model of spherical perturbations.
The details of this model will be presented elsewhere
(Mo et al. in preparation). The general idea is that
the physical radius (at a given redshift) of a
spherically symmetric perturbation is
uniquely determined by the comoving
radius $R_0$ and the extrapolated overdensity $\delta _0$
(the upper limit in the integration ensures this),
and that one can carry out the integration in Eq.(14)
for regions with the same {\it physical} radius.

In Fig.8a we plot the model prediction for
the ratio between the correlation
function of the halos of LSB galaxies and
that of average galactic halos.
Here, as discussed above,
we identify LSB galaxies
with galactic-sized halos which
have not been incorporated into larger structures containing
other galaxies by the present epoch.
For a moderate bias (e.g. $b=1.5$), the abundance
of galactic-sized halos (e.g. with circular velocities
$V_c =200\kms$) peaks at $z\sim 3$; the peak redshift
is $z\sim 1$ for $b=2.5$ (see White and Frenk 1991).
So, as illustration, we take
$z=2$ as the formation epoch of average halos of galaxies.
With these assumptions, several examples
are presented.
The two dashed curves show the results
for models with $b=1$ (short-dashed) and $b=2$ (long-dashed),
and in which both ``LSB galaxies'' and ``average galaxies''
correspond to halos with a circular velocity
$V_c=200\kms$. Since for a given circular velocity the halo
mass goes with redshift as $(1+z)^{-3/2}$, the mass
of the halos of ``LSB galaxies'' in the above models is about
five times that of ``average galaxies'' at formation.
The dot-dashed curve shows a model with $b=1.5$, and in which
halos for both LSB and average galaxies have a mass
$10^{12}M_ \odot $ at formation.
To compare model predictions
and our results, we also have to assume
that the correlation functions of our trace
galaxies differ from the mass correlation function
only by a {\it constant} bias factor which is cancelled
out in the ratios shown in the figure.

It is gratifying that
the simple models give the same trend
as the data: the ratios remain relatively flat
for separations $r>2\mpc$, and decrease with
decreasing $r$ for smaller separations.
It is also interesting
to note that the break scales and the amplitudes
of the ratios at large separations predicted by the models
are comparable to those in the data.
The agreement may be fortuitous,
considering the uncertainties both in assigning galaxies
to halos and in the evolution of the correlation functions,
but no fine tuning of parameters is necessary to reproduce the
observational trends.
The scale of the break at $\sim 2\mpc$
corresponds perhaps to groups rather than to
individual galaxies, so that tidal interactions between
group members may enhance star formation and hence surface
brightness if the group scale structure has had time to
collapse (c.f.\ Lacey \& Silk 1991).  LSB halos will by
definition not be contained within such structures, and so
would not be expected to follow the same trends on small
scales. The merging of halos
may be more significant in high density regions, which would affect
the correlation functions for normal galaxies on small scales.
However, this would reduce the correlation on small scales
for normal galaxies rather than enhance it.
Our results suggest, therefore, that the observed deficit
of pairs of LSB galaxies on small scales, and the low
amplitude of the correlation function may both be due to the fact
that LSB galaxies are formed in halos lacking close interactions
with other halos.

According to our assumption,
the relative amplitudes
of the cross-correlation functions at large separations
(see Eq.6c) can be considered to be
the relative bias factor ($b$)
of LSB galaxies with respect to that of normal galaxies.
We therefore have
$b_{\rm LSB}:b_{\rm CfA}\approx 0.41$ and
$b_{\rm LSB}:b_{\rm IRAS}\approx 0.55$.
If normal galaxies are formed with a bias factor
of 1.5 to 2, as is required to match the observed
large-scale clustering and motions of galaxies
(see e.g. Jing et al. 1993 for a summary), the bias
factor for LSB galaxies on large scales in our sample is about 1
or smaller.

As discussed above, the model we are considering
predicts a lower value of the bias
factor for the halos of LSB galaxies (Fig.8a).
This is because halos with larger overdensities
are more likely to be in high density
regions and be incorporated into bigger halos.
For a Gaussian density field, one can form
a joint probability for the (linear) overdensity
$\delta _g$ of a galactic-sized region (with radius
$R_g$ and mass $M_g$) and the overdensity $\delta _c$
of a larger region (with mass $M_c$) surrounding it:
$p(\nu _c, \nu _g) d\nu _c d\nu _g$ (see Appendix E
in BBKS).
One can therefore calculate the mean values of
$\nu _g$ (for a give $M_g$) for average galactic
halos, and for
halos which have not been incorporated
into halos with mass $> M_c$. Figure 10 shows the ratios
${\bar \delta'_{\rm g}}/ {\bar \delta _{\rm g}}$,
where
${\bar \delta_{\rm g}}$ is the mean value of overdensities
for average halos of galactic size, and
${\bar \delta'_{\rm g}}$ is that for halos which satisfy
the `no merging' condition.
The results shown are for the CDM spectrum (dashed curves), and
for a power spectrum
$P(k) \propto k^{-1.3}$ (solid curves), for different values
of $M_c$ and $b$.
The biased model (i.e., $b=2$ for {\it normal\/} galaxies)
gives larger ratios than the unbiased
one because the structure in the former case is less
evolved, and there are, therefore, more high density
regions that have not yet been incorporated into large
collapsed systems. A larger $M_c$ also gives a
larger ratio, as expected. The ratios increase with
$R_{\rm g}$, the linear scale for galactic halos,
because, in a hierarchical clustering scenario, larger
halos form later and hence have less chance to be
incorporated into larger systems. It is encouraging
that the ratios, predicted by models with reasonable
values of $R_{\rm g}$ (e.g. $=0.5 \mpc$), are comparable to the ratios
derived from the correlation properties of the LSB galaxy sample
and inferred from the physical properties of LSB galaxies
(i.e.\ ${\bar \delta'_{\rm g}}/ {\bar \delta _{\rm g}} \simlt 1/2$;
McGaugh 1992, McGaugh et al. 1993).
That the current simple model
adequately reproduces the observational trends lends support to the
notion that galaxies did indeed form from
by gravitational instability from small fluctuations in the primordial
density field.

Given the significant environmental dependence on surface brightness
detected here, and the lack of such a dependence on luminosity
(Hamilton 1988, Alimi et al. 1988, B\"orner \& Mo 1990) or circular
velocity (Mo \& Lahav 1993), it is worth noting that
it is the height of fluctuations (or equivalently, collapse time)
which should be correlated with environment.
Indeed, as shown in Fig.8a, halos formed at $z=2$
can be more strongly correlated with mass than halos
at $z=0$, even though they have smaller masses
or circular velocities.
The height of fluctuations is more directly
related to surface brightness than to total
luminosity or mass (McGaugh 1992). So our results
suggest that a strong luminosity segregation
in galaxy clustering is not a necessary consequence
of biased galaxy formation, unless
the effect of surface brightness is also taken into
account.
It is often
stated that LSB galaxies should ``fill the voids,'' but this
is misleading.  They should have lower correlation amplitudes,
but they should still be correlated.  If they form from $\sim 1 \sigma$
fluctuations
and avoid interations with other galaxies,
as seems to be the case,
then they should trace structure on scales larger than a few
megaparsecs, and possibly in an unbiased way.

\bigskip
\leftline {\bold 5.2. Model predictions}
\smallskip

Since our simple model adequately reproduces
the correlation function of LSB galaxies,
it is interesting to see what predictions it makes
for other properties of LSB galaxies.
Because LSB galaxies correspond to halos with lower
initial density in our model, they form late and
so we predict that LSB galaxies should be relatively young.
Observationally, this appears to be the case, as LSB galaxies
are unevolved (Bothun et. al. 1990, McGaugh 1993) and have
stellar populations with a young mean age
(McGaugh \& Bothun 1993, van~der~Hulst \& de~Blok 1993).
The relative formation epoch
goes roughly as $(1+z_{\rm LSB}) /(1+ z_{\rm average})
\propto {\bar \delta'_{\rm g}}/ {\bar \delta _{\rm g}}$.
If the average halos have a mean formation epoch
$z_{\rm average} \approx 2$, then the average formation epoch
for halos of LSB galaxies is about $z_{\rm LSB} \approx 0.5$.

An important issue is the total mass fraction contained
in LSB galaxies.  From Eq.~(12a), one can calculate the
mass fraction in LSB halos in a given
mass (or circular velocity) range.
The result is shown in Fig.11.
Models with higher bias parameters give higher fractions,
because a higher bias parameter corresponds to lower
amplitudes of the initial density fluctuations.
However, the fraction depends on the
maximum circular velocity
(or mass and radius) for galaxy formation, $V_{\rm u}$, and
the degree of isolation required to prevent the enhancement
of surface brightness by tidal interaction.  The mass fraction
of LSB galaxy halos increases with $V_{\rm u}$ because larger halos
form later and so have less chance to be incorporated into larger
systems (and are thus considered to be LSB galaxy halos in
our simple model).  Choosing the correct $V_{\rm u}$
requires one to distinguish between the halos which will
form groups of galaxies and those which will form single, giant
galaxies.  This in turn requires knowledge of the details of
the merging history of halos (see subsection 5.1), and $V_{\rm u}$
may well be dependent upon environment.  In this context, it is
worth noting that very massive but very LSB
galaxies like Malin~1 (Bothun et al. 1987) do exist.
These would carry a large fraction of the mass of the universe
within the framework of the current model.  Even for modest
$V_{\rm u} \approx 250\mpc$ (not large enough to accommodate
the existence of Malin~1), the model predicts that up to 30\%
of the mass can be in LSB galaxy halos.
However, this fraction could
be rather lower if weak interactions are sufficient
to induce significant star formation.  The large scale ($\sim 2\mpc$)
of the break in the LSB correlation function would seem to
argue against this, but clearly the simple model does not
provide a robust prediction of the mass fraction of LSB galaxies.
It is suggestive that this could be significant, though.

\bigskip
\centerline{\Bold 6. Summary}
\bigskip

We have investigated the clustering
properties of LSB galaxies in the galaxy density field.
We have used a cross correlation technique to
determine the cross-correlation functions between LSB galaxies
and (trace) galaxies in complete (CfA and IRAS) samples,
which enables us to compare directly the amplitudes and shapes
of the correlation functions for LSB galaxies to
those for `normal' galaxies. Our main results are
summarized as follows:

1)
For pair separations $r\gs 2\mpc$,
the cross-correlation functions between the LSB galaxies
and the trace galaxies,
$\xiab (r)$, have approximately the same shape, with power
index $\gamma\approx 1.7$, as
the auto-correlation functions for
CfA and IRAS galaxies.
The amplitudes ($A$) of the cross-correlation functions
are lower, with
$A_{\rm LSB-CfA}:A_{\rm CfA-CfA}\approx 0.41$ and
$A_{\rm LSB-IRAS}:A_{\rm IRAS-IRAS}\approx 0.55$.
Hence LSB galaxies are imbedded in the same large scale
structures as normal galaxies, but are significantly
less strongly clustered.  This is the expectation
for galaxies which form at the maxima in Gaussian random
fields when LSB galaxies correspond to lower initial densities.
This lends support to hierarchical models with
Gaussian initial conditions, and
suggests that LSB galaxies may be unbiased tracers
of the mass density.

2)
For $r\ls 2\mpc$, the cross-correlation functions are significantly
lower than expected from the extrapolation
of $\xiab$ on larger scales, suggesting that the formation of
LSB galaxies (or their survival as such) may be affected by
interaction with neighboring galaxies.

3)
We have shown that a simple hierarchical model, in which LSB
galaxies form only in halos lacking close interactions
with other halos, successfully reproduces both the observed deficit of
LSB galaxy companions and
the low amplitude of the correlation function for
LSB galaxies.

4)
This model predicts that LSB galaxies are, on average,
younger than normal galaxies, consistent with observations.
Such a model also suggests that a significant fraction of the
mass in the universe is associated with LSB galaxies.
Many such galaxies
may still remain undetected because of the selection effects
against low surface brightness objects.

\bigskip
\leftline {\bold Ackowledgements}
\smallskip
We thank Simon White and the referee, Will Saunders, for
helpful comments.
HJM and SSM acknowledge support from SERC Postdoctoral Fellowships.
\vfill
\eject

\centerline {\Bold References}
\smallskip


\hyphenation {non-re-la-ti-vi-stic}
%
\ref Alimi, J.~M., Valls-Gabaud, D., \& Blanchard, A. 1988, A\&A, 206, L11
\ref Bardeen J., Bond, J.R., Kaiser, N. \& Szalay, A.S. 1986,
{ApJ},{304}, 15 (BBKS)
\ref Barrow J.D., Bhavsar S.P., Sonoda D.H. 1984, MNRAS,
{210}, 19
\ref Bond, J.~R., Cole, S., Efstathiou, G., Kaiser, N. 1991,
ApJ, 379, 440
\ref B\"orner, G., \& Mo, H. J.
1990, A\&A, 227, 324
\ref B\"orner, G., Mo, H. J., \& Zhou, Y.Y., 1989, A\&A, 221, 191
\ref Bothun, G.~D., Beers, T.~C., Mould, J.~R., \& Huchra, J.~P.
1986, ApJ, 510
\ref Bothun, G. D., Impey, C. D., Malin, D. F., \& Mould, J. R.
1987, AJ, 94, 23
\ref Bothun, G.~D., Schombert, J.~M., Impey, C.~D., \& Schneider, S.~E.
1990, ApJ, 360, 427
\ref Bothun, G.~D., Schombert, J.~M., Impey, C.~D., Sprayberr, D. \&
McGaugh, S.~S. 1993, AJ, 106, 548
\ref Bower, R. J. 1991, MNRAS, 248, 332
\ref Cornell, M., Aaronson, M., Bothun, G. D., \& Mould, J. R. 1987,
ApJS, 64, 507
\ref Davis, M. \& Djorgovski, S. 1985, ApJ, 299, 15
\ref Davis, M. \& Geller, M.J. 1976, ApJ, 208, 13
\ref Davis, M. \& Huchra, J. 1982, ApJ, 254, 437
\ref Davis M. \& Peebles P.J.E. 1983: {ApJ}, {267}, 465
\ref Davies, J. I., Phillipps, S., Cawson, M. G. M., Disney, M. J.,
\& Kibblewhite, E. J. 1988, MNRAS, 232, 239
\ref Dekel, A., \& Silk, J. 1986, ApJ, 303, 39
\ref Disney, M.~J. 1976, Nature, 263, 573
\ref Efstathiou, G., Ellis, R. S., \& Peterson, B. A.,
1988a, MNRAS, 232, 431
\ref Efstathiou, G., Frenk, C. S., White, S. D. M., \& Davis, M.
1988b, MNRAS, 235, 715
\ref Giovanelli R., Haynes, M. P., \&
Chincarini, G. L., 1986, ApJ, 300, 77
\ref Hamilton, A.~J.~S. 1988, ApJ, 331, L59
\ref Huchra J., Davis M., Latham D. \& Tonry J. 1983,
ApJS, 52, 89
\ref Impey, C. D. \& Bothun, G. D. 1989, ApJ, 341, 89
\ref Impey, C.~D., Irwin, M.~J., Sprayberry, D., \& Bothun, G.~D.
1993, preprint
\ref Irwin, M.~J., Davies, J.~I., Disney, M.~J., \& Phillipps, S.
1990, MNRAS, 245, 289
\ref Jing, Y. P., Mo, H. J., \& B\"orner, G. 1991, A\&A, 252, 449
\ref Jing, Y. P., Mo, H. J., B\"orner, G.
\& Fang, L. Z. 1993, A\&A, in press
\ref Kaiser, N. 1984, ApJ, 284, L9
\ref Kaiser, N. 1987, MNRAS, 227, 1
\ref Knezek, P. 1992, Ph.D. Thesis, University of Massachusetts
\ref Lacey, C., \& Silk, J. 1991, ApJ, 381, 14
\ref McGaugh, S. S. 1992, Ph.D. Thesis, Univ. of Michigan
\ref McGaugh, S. S. 1993, ApJ, in press
\ref McGaugh, S. S., Bothun, G. D. 1993, AJ, submitted
\ref McGaugh, S. S., Bothun, G. D., van der Hult, J. M., \&
Schombert, J. M. 1993, in preparation
\ref Mo, H. J., Einasto, M., Deng, Z. G. \& Xia, X. Y.
1992, MNRAS, 255, 382
\ref Mo, H.~J., Jing Y.~P. \& B\"orner G. 1992, ApJ, 392, 452
\ref Mo, H.~J., Jing Y.~P. \& B\"orner G. 1993, MNRAS, in press
\ref Mo, H.~J., \& Lahav, O. 1993, MNRAS, 261, 895
\ref Mo, H. J., Miralda-Escude, J., \& Rees, M. J., 1993,
MNRAS, in press
\ref Nilson, P. 1973, {Uppsala General Catalog of Galaxies},
Uppsala, Sweden: Societatis Scientiarum Upsaliensis (UGC)
\ref Peebles P. J. E. 1980, {The Large-Scale Structure of the Universe},
Princeton: Princeton Univ. Press
\ref Phillipps, S., \& Shanks, T. 1987, MNRAS, 229, 621
\ref Schechter, P. L. 1980, AJ, 85, 801
\ref Schneider, S.~E., Thuan, T.~X., Magri, C., \& Wadiak, J.
1990, ApJS, 72, 245
\ref Schombert, J. M., \& Bothun, G. D. 1988, AJ, 95, 1389
\ref Schombert, J.~M., Bothun, G.~D., Impey, C.~D., \& Mundy, L.~G.
1990, AJ, 100, 1523
\ref Schombert, J. M., Bothun, G. D., Schneider, S. E., \&
McGaugh, S. S. 1992, AJ, 103, 1107
\ref Strauss M.A., Davis M., Yahil A. \& Huchra J.P. 1992,
ApJ, 385, 421
\ref Strauss, M. A., Yahil, A., \& Davis, M. 1991, PASP, 103, 1012
\ref Suto, Y., \& Suginohara, T., 1991, ApJL, 370, L15
\ref Thuan, T.~X., Gott, R., \& Schneider, S.~E. 1987, ApJ, 315, L93
\ref van der Hulst, J. M., \& de Blok, W. J. G. 1993, in preparation
\ref van der Hulst, J.~M., Skillman, E.~D., Smith, T.~R.,
Bothun, G.~D., McGaugh, S.~S., \& de Blok, W.~J.~G. 1993, AJ, 106, 548
\ref White S.D.M. \& Frenk C.S. 1991, ApJ, 379, 525
\ref Yahil, A., Strauss, M. A., Davis, M., \& Huchra, J. P. 1991,
ApJ, 372, 380
\ref Zaritsky, D., \& Lorrimer, S. 1992, in {The Evolution of Galaxies
and Their Environment:  Third Teton Summer School} eds.\ H.~Thronson \&
M.~Shull
\vfill
\eject

\baselineskip=15pt

\vskip3cm
{\centerline{\bf Table 1. Samples and correlation parameters}
\medskip}
\baselineskip=15pt

{\medskip\hrule}
$$\vbox {
\tabskip 0.1truecm
\halign{#\hfil&#\hfil&#\hfil&#\hfil&#\hfil&#\hfill&#\hfill&#\hfill\cr
Case & Region &$N_{\rm g}$ &$\gamma$ &$A$ &$r_0$ &$\beta$ &$r_c$\cr
     &        &            &
&$A^{(\gamma=1.65)}$ &$r_0^{(\gamma=1.65)}$ &           &$r_c^{(\beta=2)}$\cr


CfA-CfA  &POSS&665&$1.78\pm0.21$&$49.7\pm16.7$ &$5.39\pm0.95$&            &
      \cr
         &        &    &        &$36.9\pm4.7$ &$5.49\pm0.42$ &            &
      \cr
POSS-CfA &CfA&121 &$1.50\pm0.04$&$13.2\pm6.0$ &$3.53\pm1.00$ &$1.93\pm0.18$
&2.01        \cr
         &        &    &            &$16.2\pm2.6$ &$3.33\pm0.31$ &
&$1.95\pm0.13$ \cr
Spir-Spir&POSS&378&$1.69\pm0.26$&$29.4\pm13.0$&$4.50\pm1.08$ &            &
       \cr
         &        &    &            &$27.2\pm4.8$ &$4.56\pm0.47$ &            &
           \cr
POSS-Spir&CfA&121 &$1.58\pm0.35$&$8.9\pm5.6$ &$2.49\pm0.90$&$1.51\pm0.11$&2.04
      \cr
         &        &    &            &$12.4\pm2.6$ &$2.83\pm0.34$ &
&$1.78\pm0.13$ \cr
IRAS-IRAS&POSS&388&$1.91\pm0.24$&$25.5\pm10.4$&$3.20\pm0.63$&            &
      \cr
         &        &    &            &$17.8\pm2.1$ &$3.53\pm0.23$ &            &
           \cr
POSS-IRAS&POSS&151&$1.75\pm0.28$&$12.1\pm6.0$&$2.52\pm0.66$&$1.33\pm0.36$&0.77
      \cr
         &        &    &            &$10.2\pm1.5$ &$2.52\pm0.22$ &
&$1.01\pm0.25$ \cr
}}$$
{\hrule\smallskip}
\medskip
\noindent
Note: $r_0$ and $r_c$ are in Mpc/h; the sample `Spir' contains
all late type ($T>0$) galaxies in the CfA sample.
\vfill
\eject

\centerline {\bold Figure captions}
\bigskip

\noindent{\bf Figure 1}\quad
The distributions (in right-ascention and declination)
of IRAS galaxies (upper panel), CfA galaxies (lower
panel) and POSS (LSB) galaxies.

\noindent{\bf Figure 2}\quad
The redshift histograms of the CfA and IRAS samples
(upper panel), and the LSB sample (lower panel).
The smooth curves in the upper panel are the selection
functions for the CfA (dashed curve) and the IRAS (solid)
samples.

\noindent{\bf Figure 3}\quad
The average
cross-correlation function between POSS sample
and CfA sample (bullets), compared to the auto-correlation
function of CfA sample (triangles). The error
bars are given by 100 bootstrap resamplings of the POSS
sample. The straight line
corresponds to a {\it differential}
correlation function, $\xi (r)=(6\mpc /r)^{1.7} $,
which is approximately the result of the {\it redshift
space} auto-correlation function obtained by Davis \& Peebles
(1983) for the CfA sample.

\noindent{\bf Figure 4}\quad
The {\it differential}
cross-correlation function between the POSS LSB sample
and the CfA sample.
The straight line
shows a {\it differential}
correlation function, $\xi (r)=(6 \mpc /r)^{1.7}$,
which is approximately the result of the {\it redshift
space} auto-correlation function obtained by Davis \& Peebles
(1983) for the CfA sample.

\noindent{\bf Figure 5}\quad
The average
cross-correlation function between POSS sample
and spiral galaxies (with $T>0$) in the
CfA sample (circles), compared to the auto-correlation
function of the spiral galaxies (triangles).
The crosses show the cross-correlation function
estimated by using a different selection function
for the CfA sample (see text).
The straight line
corresponds to a {\it differential}
correlation function, $\xi (r)=(6 \mpc /r)^{1.7}$,
which is approximately the result of the {\it redshift
space} auto-correlation function obtained by Davis \& Peebles
(1983) for the CfA sample.

\noindent{\bf Figure 6}\quad
The average
cross-correlation functions between the POSS sample
and the IRAS sample (bullets),
compared to the auto-correlation
function of IRAS sample (triangles). The error
bars are given by 100 bootstrap resamplings of the POSS
sample. The straight line
corresponds to a {\it differential}
correlation function, $\xi (r)=(4 \mpc /r)^{1.6}$,
which is approximately the result of the {\it redshift
space} auto-correlation function obtained by Strauss et al.
(1992) for the IRAS sample.

\noindent{\bf Figure 7}\quad
The {\it differential}
cross-correlation functions between the POSS
and IRAS samples (bullets).
The straight line
shows a {\it differential}
correlation function, $\xi (r)=(4 \mpc /r)^ {1.6} $,
which is approximately the result of the {\it redshift
space} auto-correlation function obtained by Strauss et al.
(1992) for the IRAS sample.

\noindent{\bf Figure 8}\quad
(a) The average
cross-correlation functions between the POSS LSB sample
and the CfA sample,
divided by the average auto-correlation function
of the CfA sample.
Bullets show the case (POSS-CfA)/(CfA-CfA);
squares show (POSS-Spiral)/(Spiral-Spiral).
The error
bars are given by 100 bootstrap resamplings of the POSS
sample. A horizontal line having unity value is drawn
for reference. The solid curve shows the model
Eq.(7) with $\beta=2$ and $r_c=1.95 \mpc$.
The two dashed curves show the results
for models (see subsection 5.1)
with $b=1$ (short-dashed) and $b=2$ (long-dashed),
and in which both ``LSB galaxies'' and ``average galaxies''
correspond to halos with a circular velocity
$V_c=200\kms$.
The dot-dashed curve shows a model with $b=1.5$, and in which
halos for both LSB and average galaxies have a mass
$10^{12}M_ \odot $ at formation.
(b) The average
cross-correlation functions between the POSS LSB
and IRAS samples,
divided by the average auto-correlation function
of the IRAS sample (bullets).
The error
bars are given by 100 bootstrap resamplings of the POSS
sample. The solid curve shows the model
Eq.(7) with $\beta=2$ and $r_c=1.39 \mpc$.

\noindent{\bf Figure 9}\quad
The projected
cross-correlation function between the POSS LSB sample
and the CfA sample.
The straight dotted line
has a slope $-0.7$ which corresponds to a
{\it real space}
correlation function with $\gamma =1.7$.

\noindent{\bf Figure 10}\quad
The model predictions (subsection 5.1) for the ratios
between the
average overdensities of the halos of LSB galaxies
and those of average galaxies, as a function
of the linear size of galactic halos.
The results shown are for the CDM spectrum (dashed curves), and
for a power spectrum
$P(k) \propto k^{-1.3}$ (solid curves), for different values
of the bias parameter $b$ and mass $M_c$ (the thick
curves show $M_c/M_g=2$).

\noindent{\bf Figure 11}\quad
The model predictions (subsection 5.2) for the
mass fractions in LSB halos with circular velocities
in the range from $V_l$ to $V_{\rm u}$.
The results are shown for CDM models with
bias parameters $b=1$ (dashed curves), and
$b=2$ (dot-dashed curves), and for
two values of $V_l$: $50\kms$ and $100\kms$.
These bracket the range of $V_l$ below which
a halo is likely to form a dwarf rather than
a LSB disk galaxy.
\bye